\documentclass[a4,amssymb,amsmath,aps,prl,reprint,superscriptaddress,showpacs,floatfix]{revtex4-1}
\usepackage{amssymb,amsmath}
\usepackage{color}
\usepackage{graphicx}
\usepackage{epstopdf}
\usepackage{array}
\usepackage{braket}
\usepackage{tikz}
\usepackage{pgfplots}
\usepackage{bm}
\usepackage{hyperref}
\hypersetup{colorlinks=true, citecolor=blue, urlcolor=blue, linkcolor=red}

\newcommand{\naoso}{NaOsO$_{\text{3}}$}
\newcommand{\cdoso}{Cd$_{\text{2}}$Os$_{\text{2}}$O$_{\text{7}}$}

\begin{document}
\title{Evolution of the magnetic excitations in NaOsO$_{\text{3}}$ through its metal-insulator transition}

\author{J. G. Vale}
\email{j.vale@ucl.ac.uk}
\affiliation{London Centre for Nanotechnology and Department of Physics and Astronomy, University College London, Gower Street, London, WC1E 6BT, United Kingdom}
\affiliation{Laboratory for Quantum Magnetism, \'{E}cole Polytechnique F\'ed\'erale de Lausanne (EPFL), CH-1015, Switzerland}

\author{S. Calder}
\email{caldersa@ornl.gov}
\affiliation{Quantum Condensed Matter Division, Oak Ridge National Laboratory, Oak Ridge, Tennessee 37831, USA}

\author{C. Donnerer}
\affiliation{London Centre for Nanotechnology and Department of Physics and Astronomy, University College London, Gower Street, London, WC1E 6BT, United Kingdom}

\author{D. Pincini}
\affiliation{London Centre for Nanotechnology and Department of Physics and Astronomy, University College London, Gower Street, London, WC1E 6BT, United Kingdom}
\affiliation{Diamond Light Source, Harwell Science and Innovation Campus, Didcot, Oxfordshire, OX11 0DE, United Kingdom}

\author{Y. G. Shi}
\affiliation{Beijing National Laboratory for Condensed Matter Physics and Institute of Physics, Chinese Academy of Sciences, Beijing 100190, China}
\affiliation{Research Center for Functional Materials, National Institute for Materials Science, 1-1 Namiki, Tsukuba, Ibaraki 305-0044, Japan}

\author{Y. Tsujimoto}
\affiliation{Research Center for Functional Materials, National Institute for Materials Science, 1-1 Namiki, Tsukuba, Ibaraki 305-0044, Japan}

\author{K. Yamaura}
\affiliation{Research Center for Functional Materials, National Institute for Materials Science, 1-1 Namiki, Tsukuba, Ibaraki 305-0044, Japan}
\affiliation{Graduate School of Chemical Sciences and Engineering, Hokkaido University, North 10 West 8, Kita-ku, Sapporo, Hokkaido 060-0810, Japan}

\author{M. Moretti Sala}
\affiliation{ESRF, The European Synchrotron, 71 Avenue des Martyrs, 38043 Grenoble, France}

\author{J. van den Brink}
\affiliation{Institute for Theoretical Solid State Physics, IFW Dresden, D01171 Dresden, Germany}

\author{A. D. Christianson}
\affiliation{Quantum Condensed Matter Division, Oak Ridge National Laboratory, Oak Ridge, Tennessee 37831, USA}
\affiliation{Department of Physics and Astronomy, University of Tennessee, Knoxville, TN 37996, USA}

\author{D. F. McMorrow}
\affiliation{London Centre for Nanotechnology and Department of Physics and Astronomy, University College London, Gower Street, London, WC1E 6BT, United Kingdom}


\begin{abstract}
The temperature dependence of the excitation spectrum in \naoso\ through its metal-to-insulator transition (MIT) at 410 K has been investigated using resonant inelastic X-ray scattering (RIXS) at the Os L$_{\text{3}}$ edge. High resolution ($\Delta E \sim \text{56~meV}$) measurements show that the well-defined, low energy magnons in the insulating state weaken and dampen upon approaching the metallic state. Concomitantly, a broad continuum of excitations develops which is well described by the magnetic fluctuations of a nearly antiferromagnetic Fermi liquid. By revealing the continuous evolution of the magnetic quasiparticle spectrum as it changes its character from itinerant to localized, our results provide unprecedented insight into the nature of the MIT in \naoso.\footnote{This work was prepared as a joint submission with Physical Review B: please see Ref.~\onlinecite{vale2018_prb} and references therein.}
\end{abstract}
\maketitle

The nature of the MIT in transition metal oxides (TMOs) is of enduring interest as it represents a spectacular manifestation of competing interactions, and their effects on the most fundamental transport property of materials. Recently, considerable attention has focussed on the nature of the MIT in 5d TMOs, which have the additional ingredients  of strong spin-orbit coupling (SOC) and a lower degree of localization relative to their 3d counterparts \cite{witczak-krempa2014, rau2016}. 
New phenomenology has been revealed by experiment, including a number of MITs which are intimately entwined with the onset of magnetic order. Notably these MITs do not appear to be associated with any spontaneous lattice symmetry breaking, placing them outside of the conventional Mott-Hubbard paradigm. A key and yet unsolved issue for such systems is to fully determine the nature of the magnetic quasiparticle spectrum as the electronic character evolves from itinerant to localized through the MIT.

\begin{figure}[t]
\centering
\includegraphics{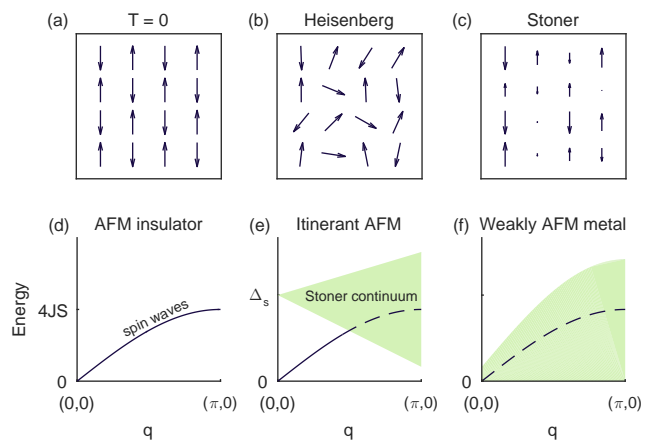}
\caption{Color online. Top panels: Schematic of the orientation of antiferromagnetically (AFM) interacting magnetic moments on a square lattice for $T=\text{0}$, and finite temperatures in the Heisenberg or Stoner limits. Bottom panels: Excitation spectra in the Heisenberg limit, for an itinerant AFM (with critical energy scale $\Delta_{\text{s}}$), and for a metal with weak AFM correlations. Dashed lines indicate damped excitations.}
\label{Stoner}
\end{figure}

The classification of magnetic interactions between full itineracy and localization is usually described in one of two limits \cite{moriya1985}.
In the local moment (Heisenberg) limit, magnetism in insulators is assumed to arise from unpaired electrons within an atomic picture.
At some temperature $T$, the orientation of the magnetic moments varies, but their magnitude remains fixed at the $T=\text{0}$ value.
Consequently only the transverse component of the local spin-density fluctuation (LSF) $\chi^{\pm}(q,\omega)$ is important. This gives rise to collective spin wave excitations and is generally applicable to magnetic insulators.
At the other extreme (itinerant or Stoner limit), the electrons and spin fluctuations are extended in real space; with thermodynamic properties of the system governed by intraband electron-hole pair interactions, which have a degree of collective behaviour. At some temperature $T$, the orientation of the magnetic moments remains fixed, but their magnitude is reduced from the $T=\text{0}$ value. In this limit, the longitudinal component $\chi^{zz}(q,\omega)$ and the temperature dependence of the LSF are important.
Between these two limits, one typically observes Landau damping of spin waves by intraband particle-hole excitations. Damping is more prevalent at large wavevectors $q$ since there are a greater number of available states for the corresponding magnon to decay into. Such an effect has been observed in a number of materials, notably doped La$_{\text{2-x}}$Sr$_{\text{x}}$CuO$_{\text{4}}$ \cite{letacon2011, dean2013, monney2016}, and a subset of the iron pnictides \cite{diallo2009, zhao2009, zhou2013, wang2013, leong2014, dai2015}. A summary of the behaviour in the respective limits is shown schematically in Fig.~\ref{Stoner}.

Here we focus our attention upon materials which undergo continuous MITs with temperature, that are concomitant with the onset of long-ranged, commensurate antiferromagnetic order. Moreover, the crystallographic space group remains constant through the MIT.
Examples include some of the 5$d^5$ pyrochlore iridates  R$_{\text{2}}$Ir$_{\text{2}}$O$_{\text{7}}$ (R = Ln$^{\text{3+}}$) \cite{matsuhira2011, nakayama2016}, plus the 5$d^3$ osmates \cdoso\ \cite{mandrus2001, padilla2002, yamaura2012}, and \naoso. The latter material has been proposed to be an example of a Slater insulator \cite{shi2009, calder2012, du2012, jung2013, lovecchio2013}, in which the formation of antiferromagnetic order itself at $T_{\text{MI}}=\text{410~K}$ drives the formation of an insulating gap below the N\'{e}el temperature. This interpretation, however, is not universally accepted \cite{bongjaekim2016}.
Nevertheless, together with significant spin-phonon coupling \cite{calder2015_naoso3}, one observes an unprecedented connection between the magnetic, electronic, structural, and phonon degrees of freedom in \naoso. 

One would thus expect the electronic and magnetic excitations to evolve correspondingly through the MIT. Optical conductivity measurements on \naoso\ reveal a continuous opening of the electronic gap with decreasing temperature $\left\lbrace\Delta_g (0) = \text{102(3)~meV}\right\rbrace$ \cite{lovecchio2013}, and an MIT in which electronic correlations play a limited role. This is consistent with a  Slater picture in which interactions are mean-field like.
Meanwhile previous RIXS measurements showed well-defined and strongly gapped ($\sim\!\text{50~meV}$) dispersive spin-wave excitations at 300~K \cite{calder2017_naoso3}.  This was found to be consistent with an anisotropic nearest-neighbour Heisenberg picture for the magnetic Hamiltonian, and is suggestive of localized magnetic moments. 
The question remains whether either the spin or electronic excitations remain coherent through the MIT, and if there is any evidence of coupling to any of the other relevant degrees of freedom present in the system.

In this manuscript (and Ref.~\onlinecite{vale2018_prb}), we establish that there is a continuous progression from itinerant to localized behavior through the MIT in \naoso. This is revealed by a significant renormalization of the magnetic quasi-particle spectral weight over large ranges of momentum and energy.
In particular, the presence of correlations in the metallic state immediately leads to a deviation from mean-field behavior, and hence, true Slater phenomenology. 

Our experiments relied on exploiting the unique ability of RIXS to provide momentum-resolved 
sensitivity to the excitations of the orbital, electronic and magnetic degrees of freedom.
By providing data on an experimental test case, in which the magnetic quasiparticle spectrum can be tuned simply by varying the temperature, our work in turn helps extend the utility of RIXS, which has hitherto been best understood in the localized limit \cite{haverkort2010, jia2014, kim_khaliullin2017}.

RIXS measurements were performed at the Os $L_3$ edge ($E = \text{10.877~keV}$, $2p\rightarrow 5d$) on the ID20 spectrometer at the ESRF, Grenoble. Preliminary measurements were performed at 9-ID-B, Advanced Photon Source. The energy resolution of the spectrometer was estimated to be $\Delta E = \text{56~meV}$, based upon scattering from transparent adhesive tape. The incident energy used equates to 3~eV below the maximum of the white line obtained from x-ray absorption spectroscopy (XAS). Further experimental details are provided in Ref.~\onlinecite{vale2018_prb}.

Our initial measurements focussed upon the high energy orbital excitations from the ground state (intra-$t_{2g}$ and $t_{2g}\rightarrow e_g$), which were consistent with those observed in Ref.~\onlinecite{calder2017_naoso3}. We were unable to determine any significant temperature dependence of these broad excitations. This corroborates the supposition that the MIT is not driven by local structural distortion, which would manifest in significant variations of the local crystal field parameters. 

The main focus of this paper is the low-energy excitation spectrum below 0.5~eV. RIXS spectra were collected at three different momentum transfers as a function of temperature: $\Gamma$ $(\text{4.95},\,\text{2.95},\,\text{3.95})$, $\Gamma$--$Y$ $(\text{5},\,\text{2.75},\,\text{4})$, and $Y$ $(\text{5},\,\text{2.5},\,\text{4})$. This reflects a progression from the Brillouin zone centre to the zone boundary. Note that \naoso\ exhibits $\bm{Q}_{\mathrm{AFM}}=0$ order, that is, the structural and magnetic lattices are commensurate with one another. Hence the point near $\Gamma$ was chosen in order to avoid the weak magnetic Bragg peak at $(\text{5},\,\text{3},\,\text{4})$. 
Four RIXS spectra were collected for each temperature and momentum transfer (30s/pt). 
These spectra were each normalized to the intensity of the intra-$t_{2g}$ excitations at 1~eV energy loss, cross-correlated to account for any temporal shift in the elastic line position, then averaged. 

\begin{figure*}
\includegraphics{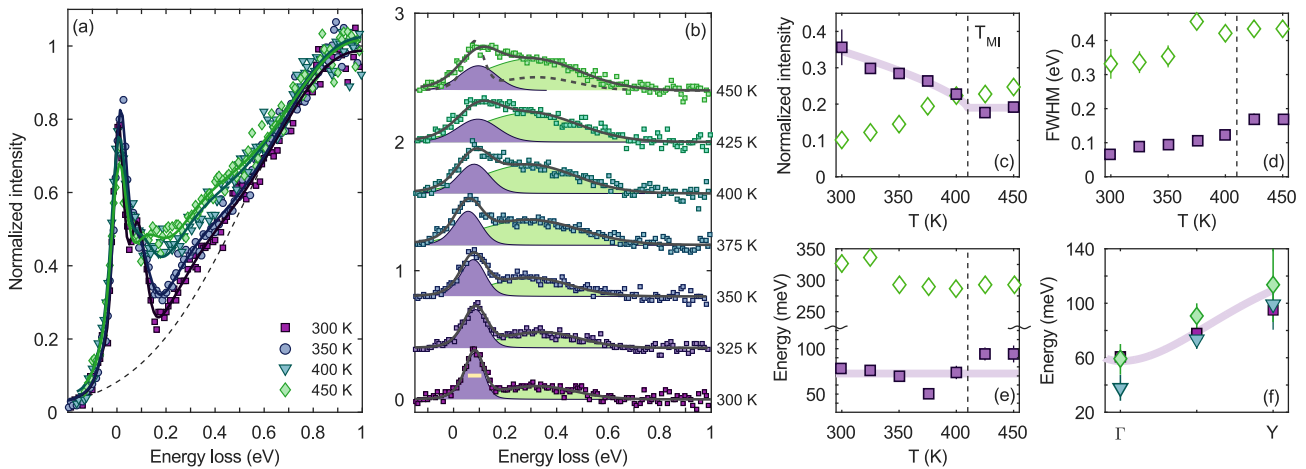}
\caption{Color online. Analysis of RIXS spectra collected at $\Gamma$--$Y$. (a): Representative low-energy RIXS spectra, each of which is normalized to the d-d excitations at 1~eV energy loss (dashed line). (b): Spectra with elastic line and d-d contributions subtracted off. Added are the best fit to the data (black solid line), and relative components of the magnon peak (purple) and high-energy continuum (green). Dashed line superimposed on 450~K plot is best fit to 300~K data for comparison. Yellow bar indicates FWHM of resolution function. (c--e): Fitted peak intensity (c), intrinsic FWHM (d), and energy (e) of the magnon peak (filled squares) and high-energy continuum (open diamonds) as a function of temperature. Solid lines are guides to the eye. (f): Energy of magnon peak as function of momentum transfer and temperature. The symbols are the same as part (a). Solid line is best fit to dispersion at 300~K as determined within Ref.~\onlinecite{calder2017_naoso3}.}
\label{gamma_y_fits}
\end{figure*}
Representative RIXS spectra (at $\Gamma$--$Y$) are plotted in Fig.~\ref{gamma_y_fits}a, for temperatures below and above the MIT \footnote{Data for other momentum transfers are presented in Ref.~\onlinecite{vale2018_prb}.}.
They are also shown with the elastic line and $d$-$d$ contributions subtracted (Figs.~\ref{gamma_y_fits}b, in order to better isolate changes to the spectra below 1~eV.
The spectrum at 300~K is in agreement with that given in Ref.~\onlinecite{calder2017_naoso3}, with a sharp dispersive peak evident at 60--100~meV energy loss attributable to a single magnon excitation.
With increasing temperature this peak progressively weakens and broadens. Strikingly, concurrent with the diminishing of the single magnon peak, there is a continuous increase in intensity between 0.1 and 0.6~eV, whilst there is no significant change in the intra-$t_{2g}$ excitations.

In order to quantify these observations further, the data were fitted with Gaussians to represent the elastic line, magnon peak, broad component centred around 300~meV, and the intra-$t_{2g}$ excitations. The fits in the low energy portion of the RIXS spectra were corrected to take the Bose factor into account. Prior to fitting, the model lineshape was convoluted with the (asymmetric) experimental resolution function \cite{vale2018_prb}.
This minimal model for the lineshape was used in order to reduce the number of free parameters in the fit, whilst allowing the relevant features of the data to be captured. The results of these fits are given in the remaining panels of Fig.~\ref{gamma_y_fits}.
Our simple model gives a good description of the experimental data, with $0.8\leq\chi^2/\nu\leq 1.3$ for all datasets.
Clearly there is a significant variation of the RIXS spectra through the MIT, as \naoso\ progresses from the localized to the itinerant limit. We briefly discuss the temperature dependence of the fitted parameters.

Firstly the peak at 80~meV appears to weaken progressively with increasing temperature for all momentum transfers (Fig.~\ref{gamma_y_fits}c). This abates at $T_{\text{N}}$, with some residual intensity remaining all the way to 450~K; and confirms its previous assignment as a magnon peak \cite{calder2017_naoso3}. 
Secondly the energy of the magnon peak appears to weakly vary with temperature (Fig.~\ref{gamma_y_fits}e). However, the magnon dispersion at 450~K appears remarkably similar to that collected at 300~K (within experimental uncertainty, Fig.~\ref{gamma_y_fits}f). 
We note that the previously reported value for the effective uniaxial anisotropy ($\Gamma=\text{1.4(1)~meV}$) is likely to be an overestimate of the true magnitude. This is due to the effect of the finite momentum resolution of the RIXS spectrometer, which was not taken into account within the minimal model presented in Ref.~\onlinecite{calder2017_naoso3}. 
Combined with the finite energy resolution, this means that any temperature dependence of the spin gap is likely to be washed out and difficult to observe.

Finally the width of the magnon peak increases as a function of increasing temperature for all momentum transfers (Fig.~\ref{gamma_y_fits}d). In Ref.~\onlinecite{calder2017_naoso3}, it was proposed that the well-defined, resolution limited spin wave excitations at 300~K were evidence of localized (Heisenberg) behavior.
Yet with the aid of our new fitting model, we observe experimentally that the magnons have an intrinsic FWHM of 30~meV at $\Gamma$, increasing to 70~meV at larger momentum transfers \cite{vale2018_prb}. Some of this damping may arise from processes known to apply in the localized limit (four-magnon, magnon-phonon coupling).
There is, however, the omnipresent continuous MIT at 410~K. Previous bulk measurements have shown that the charge gap $\Delta_g \sim \text{80~meV}$ at 300~K \cite{shi2009}, with the optical gap having a similar magnitude \cite{lovecchio2013}. Within a weak coupling theory, collective antiferromagnetic spin wave excitations are expected to merge into a Stoner continuum above a critical energy $\Delta_s$ (Fig.~\ref{Stoner}e). By using theoretical expressions given within Ref.~\onlinecite{feddersmartin_1966}, we find that $\Delta_s \sim \text{60~meV}$ for \naoso\ \cite{vale2018_prb}, implying that the magnons are (weakly) Landau damped for all wavevectors.

Here we focus on the characteristics of the RIXS spectrum in the metallic phase at 450 K, and what the magnetic quasiparticle spectrum reveals about the nature of the  electronic state of the metallic phase.
(A detailed discussion of the low-energy magnetic excitations deep in the insulating antiferromagnetic phase
can be found in Ref. \onlinecite{calder2017_naoso3}.)
\begin{figure}[t!]
\centering
\includegraphics{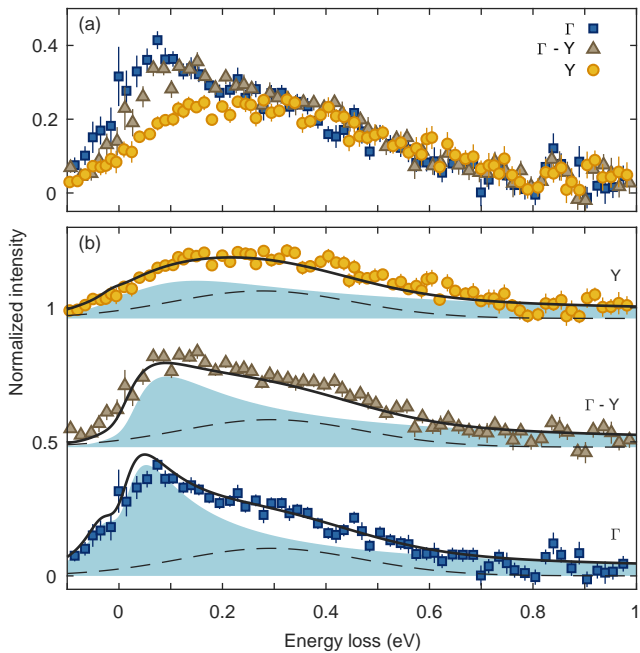}
\caption{Color online. Paramagnetic spin fluctuations (SF) in \naoso\ at 450~K. 
(a): Experimental data collected at various crystal momenta. For clarity the elastic line and high-energy intra-$t_{2g}$ excitations have been subtracted. (b): Comparison of WAFL model (solid line) with experimental data. Filled area indicates paramagnetic SF calculated using Equation \ref{dynamicsusceptibility}, with  $\gamma=0.02~\mathrm{meV}^{-1}$ and $\xi/a_0\sim 1$. Dashed line is a momentum-independent high-energy contribution, which likely results from inter-band particle-hole excitations.}
\label{PM_fluctuations}
\end{figure}

The RIXS spectra are plotted in Fig.~\ref{PM_fluctuations}a; the elastic and intra-$t_{2g}$ contributions have been subtracted in order to highlight the salient features. A broad asymmetric feature can be observed, the shape and amplitude of which varies with momentum transfer. Meanwhile the high-energy portion ($>\text{0.3~eV}$) appears momentum \emph{independent}. The simple underdamped fitting model used in Fig.~\ref{gamma_y_fits} leads to somewhat ambiguous results above $T_N$; this is especially the case for higher momentum transfers where the excitations are weaker \cite{vale2018_prb}.
We therefore considered models which more naturally describe spin fluctuations in the metallic phase, noting that previous bulk measurements \cite{shi2009} suggest \naoso\ exhibits paramagnetic Fermi liquid behavior above the N\'{e}el temperature. 

Specifically, we used self-consistent renormalization (SCR) theory appropriate for a weakly antiferromagnetic Fermi liquid (WAFL) \cite{moriya1985}. 
In this picture, magnetic order is destroyed by thermal excitations of long wavelength spin fluctuations. Yet even though long-ranged order disappears at $T_{\text{N}}$, there may be persistent short-ranged antiferromagnetic correlations, which are characterized by some correlation length $\xi$. These localized clusters of antiferromagnetic order are able to diffuse through the crystal lattice, giving rise to incoherent spin excitations. 

For simplicity we assume that the fluctuations are spatially isotropic, and consider a pseudo-cubic unit cell, with a lattice constant $a_0=3.80~\text{\AA}$ given by the Os-Os distance.
Following the approach of Inosov and Tucker \cite{inosov2010, tucker2014}, the imaginary part of the dynamic susceptibility $\chi^{''}(\mathbf{Q}, E)$ is given by:
\begin{equation}\label{dynamicsusceptibility}
\chi^{''}(\mathbf{Q},E) \propto \frac{\chi_0\Gamma E}{E^2 + \Gamma^2\left[1+\xi^2 (\mathbf{Q}-\mathbf{Q}_{\text{AFM}})^2 \right]^2}.
\end{equation}
In this expression the spin relaxation rate $\Gamma$ is defined through $\Gamma\equiv a_0^2/\gamma\xi^2$, where $\gamma$ denotes the damping coefficient arising from spin decay into particle-hole excitations (related to the electronic band structure), and $\xi$ is the spin-spin correlation length. Meanwhile $\chi_0$ is the staggered susceptibility at $\bm{Q}_{\text{AFM}}=0$ \cite{diallo2010}. All of these parameters are in principle dependent upon temperature.
For completeness the anti-Stokes (energy gain) process has also been included; this is a factor of $\exp{(-\hbar\omega/kT)}$ weaker than the equivalent Stokes (energy loss) process.

There is remarkable agreement between this model for the spin fluctuations and the data (Fig.~\ref{PM_fluctuations}b), with the best global fit to the data using $\gamma\sim 0.02~\text{meV}^{-1}$ and $\xi/a_0\sim 1$. The magnitude of $\gamma$ is similar to that found in overdoped Ba(Fe$_{\text{1-x}}$Co$_{\text{x}}$)$_{\text{2}}$As$_{\text{2}}$ \cite{tucker2014}, indicative of itinerant behavior. Moreover, the fact that $\xi$ is on the order of $a_0$ implies that the spin fluctuations at 450~K are short-ranged, and that the system is close to the hydrodynamic limit.

We make the following comments. 
The experimental RIXS cross-section includes a number of additional contributions, which include momentum-dependent absorption and polarization effects. This explains the use of a proportionality in Eqn.~\ref{dynamicsusceptibility}, in order to model these contributions phenomenologically. 
Secondly quantitative agreement with the data was improved through the inclusion of an additional scattering component centered at 0.3~eV (dashed line in Fig.~\ref{PM_fluctuations}b). This component was constrained to be momentum-independent, in order to facilitate convergence. 
The nature of this component is unclear directly from the data. This is because RIXS is sensitive to both electronic and magnetic interactions, and it is difficult to distinguish them \emph{a priori} without further information. We posit that it results primarily from inter-band particle-hole excitations, based on calculations presented within Ref.~\onlinecite{vale2018_prb}.

Finally, the fits presented in Fig.~\ref{PM_fluctuations}b explicitly use the same scale factor for all momentum transfers, in order to remove any ambiguity related to the proportionality in Eqn.~\ref{dynamicsusceptibility}. The fact that the WAFL model describes the experimental data so well at 450 K, without the use of arbitrary scale factors, is further evidence that this better reflects the behavior in the metallic phase as opposed to the phenomenological underdamped model plotted in Fig.~\ref{gamma_y_fits} \footnote{Further improvements in the quantitative agreement with the experimental data could likely be made by considering anisotropic spin fluctuations, or the effect of finite momentum resolution of the spectrometer. However, there are insufficient data to perform this robustly.}. 
\\

In this manuscript, we provide evidence that the continuous MIT observed in \naoso\ extends to a continuous evolution of the magnetic quasiparticle spectrum.
The magnetic excitations at 300~K appear to be weakly Landau damped at all wavevectors, as a consequence of the emergent charge gap below the MIT.
Meanwhile at 450~K, these excitations become diffusive, and can be successfully modelled within a weakly antiferromagnetic Fermi liquid picture. Such an approach naturally requires the presence of correlations in the high temperature phase, which leads to a departure from mean-field behavior required for a pure Slater MIT. 
Therefore we conclude that \naoso\ lies proximate to, but not within, the Slater limit.

A number of similarities can be drawn between the behavior of the magnetic excitations in \naoso, and those observed for the cuprates and pnictides as a function of carrier doping. The distinction in \naoso\ is that the continuous evolution of the magnetic quasiparticle spectra occurs within a single system, through tuning of temperature alone. This is in contrast with the cuprates and pnictides, where measurements have to be performed on multiple systems with different doping levels, slightly different experimental conditions, and so forth.
Hence \naoso\ is an ideal system for testing theoretical predictions of the RIXS response as a function of the strength of electronic interaction. The results presented here (and in Ref.~\onlinecite{vale2018_prb}) suggest a correspondence between the dynamic structure factor $S(\bm{Q},\omega)$ and the RIXS cross-section in the itinerant limit.

\acknowledgments{J.~G.~V. thanks University College London (UCL) and \'{E}cole Polytechnique F\'{e}d\'{e}rale de Lausanne (EPFL) for financial support through a UCL Impact award, and useful discussions with B. J. Blackburn, A. Princep, and E. V\"{a}is\"{a}nen. Work at UCL was supported by the EPSRC (grants EP/N027671/1, EP/N034872/1).  This research used resources at the High Flux Isotope Reactor and Spallation Neutron Source, DOE Office of Science User Facilities operated by the Oak Ridge National Laboratory. K.Y. thanks financial support from JSPS KAKENHI (15K14133 and 16H04501). All data created during this research are openly available from the UCL Discovery data archive at \url{http://dx.doi.org/10.14324/000.ds.1550187}.}


\begin{thebibliography}{35}%
\makeatletter
\providecommand \@ifxundefined [1]{%
 \@ifx{#1\undefined}
}%
\providecommand \@ifnum [1]{%
 \ifnum #1\expandafter \@firstoftwo
 \else \expandafter \@secondoftwo
 \fi
}%
\providecommand \@ifx [1]{%
 \ifx #1\expandafter \@firstoftwo
 \else \expandafter \@secondoftwo
 \fi
}%
\providecommand \natexlab [1]{#1}%
\providecommand \enquote  [1]{``#1''}%
\providecommand \bibnamefont  [1]{#1}%
\providecommand \bibfnamefont [1]{#1}%
\providecommand \citenamefont [1]{#1}%
\providecommand \href@noop [0]{\@secondoftwo}%
\providecommand \href [0]{\begingroup \@sanitize@url \@href}%
\providecommand \@href[1]{\@@startlink{#1}\@@href}%
\providecommand \@@href[1]{\endgroup#1\@@endlink}%
\providecommand \@sanitize@url [0]{\catcode `\\12\catcode `\$12\catcode
  `\&12\catcode `\#12\catcode `\^12\catcode `\_12\catcode `\%12\relax}%
\providecommand \@@startlink[1]{}%
\providecommand \@@endlink[0]{}%
\providecommand \url  [0]{\begingroup\@sanitize@url \@url }%
\providecommand \@url [1]{\endgroup\@href {#1}{\urlprefix }}%
\providecommand \urlprefix  [0]{URL }%
\providecommand \Eprint [0]{\href }%
\providecommand \doibase [0]{http://dx.doi.org/}%
\providecommand \selectlanguage [0]{\@gobble}%
\providecommand \bibinfo  [0]{\@secondoftwo}%
\providecommand \bibfield  [0]{\@secondoftwo}%
\providecommand \translation [1]{[#1]}%
\providecommand \BibitemOpen [0]{}%
\providecommand \bibitemStop [0]{}%
\providecommand \bibitemNoStop [0]{.\EOS\space}%
\providecommand \EOS [0]{\spacefactor3000\relax}%
\providecommand \BibitemShut  [1]{\csname bibitem#1\endcsname}%
\let\auto@bib@innerbib\@empty
\bibitem [{\citenamefont {Vale}\ \emph {et~al.}(2018)\citenamefont {Vale},
  \citenamefont {Calder}, \citenamefont {Donnerer}, \citenamefont {Pincini},
  \citenamefont {Shi}, \citenamefont {Tsujimoto}, \citenamefont {Yamaura},
  \citenamefont {Sala}, \citenamefont {van~den Brink}, \citenamefont
  {Christianson},\ and\ \citenamefont {McMorrow}}]{vale2018_prb}%
  \BibitemOpen
  \bibfield  {author} {\bibinfo {author} {\bibfnamefont {J.~G.}\ \bibnamefont
  {Vale}}, \bibinfo {author} {\bibfnamefont {S.}~\bibnamefont {Calder}},
  \bibinfo {author} {\bibfnamefont {C.}~\bibnamefont {Donnerer}}, \bibinfo
  {author} {\bibfnamefont {D.}~\bibnamefont {Pincini}}, \bibinfo {author}
  {\bibfnamefont {Y.~G.}\ \bibnamefont {Shi}}, \bibinfo {author} {\bibfnamefont
  {Y.}~\bibnamefont {Tsujimoto}}, \bibinfo {author} {\bibfnamefont
  {K.}~\bibnamefont {Yamaura}}, \bibinfo {author} {\bibfnamefont {M.~M.}\
  \bibnamefont {Sala}}, \bibinfo {author} {\bibfnamefont {J.}~\bibnamefont
  {van~den Brink}}, \bibinfo {author} {\bibfnamefont {A.~D.}\ \bibnamefont
  {Christianson}}, \ and\ \bibinfo {author} {\bibfnamefont {D.~F.}\
  \bibnamefont {McMorrow}},\ }\href {\doibase --} {\bibfield  {journal}
  {\bibinfo  {journal} {Phys. Rev. B}\ }\textbf {\bibinfo {volume} {--}},\
  (\bibinfo {year} {2018})},\ \bibinfo {note} {and references
  therein}\BibitemShut {NoStop}%
\bibitem [{\citenamefont {Witczak-Krempa}\ \emph {et~al.}(2014)\citenamefont
  {Witczak-Krempa}, \citenamefont {Chen}, \citenamefont {Kim},\ and\
  \citenamefont {Balents}}]{witczak-krempa2014}%
  \BibitemOpen
  \bibfield  {author} {\bibinfo {author} {\bibfnamefont {W.}~\bibnamefont
  {Witczak-Krempa}}, \bibinfo {author} {\bibfnamefont {G.}~\bibnamefont
  {Chen}}, \bibinfo {author} {\bibfnamefont {Y.~B.}\ \bibnamefont {Kim}}, \
  and\ \bibinfo {author} {\bibfnamefont {L.}~\bibnamefont {Balents}},\ }\href
  {\doibase 10.1146/annurev-conmatphys-020911-125138} {\bibfield  {journal}
  {\bibinfo  {journal} {Ann. Rev. Cond. Mat. Phys.}\ }\textbf {\bibinfo
  {volume} {5}},\ \bibinfo {pages} {57} (\bibinfo {year} {2014})}\BibitemShut
  {NoStop}%
\bibitem [{\citenamefont {Rau}\ \emph {et~al.}(2016)\citenamefont {Rau},
  \citenamefont {Lee},\ and\ \citenamefont {Kee}}]{rau2016}%
  \BibitemOpen
  \bibfield  {author} {\bibinfo {author} {\bibfnamefont {J.~G.}\ \bibnamefont
  {Rau}}, \bibinfo {author} {\bibfnamefont {E.~K.-H.}\ \bibnamefont {Lee}}, \
  and\ \bibinfo {author} {\bibfnamefont {H.-Y.}\ \bibnamefont {Kee}},\ }\href
  {\doibase 10.1146/annurev-conmatphys-031115-011319} {\bibfield  {journal}
  {\bibinfo  {journal} {Ann. Rev. Cond. Mat. Phys.}\ }\textbf {\bibinfo
  {volume} {7}},\ \bibinfo {pages} {195} (\bibinfo {year} {2016})}\BibitemShut
  {NoStop}%
\bibitem [{\citenamefont {Moriya}(1985)}]{moriya1985}%
  \BibitemOpen
  \bibfield  {author} {\bibinfo {author} {\bibfnamefont {T.}~\bibnamefont
  {Moriya}},\ }\href@noop {} {\emph {\bibinfo {title} {Spin Fluctuations in
  Itinerant Electron Magnetism}}},\ Springer Series in Solid-State Sciences\
  (\bibinfo  {publisher} {Springer-Verlag Berlin Heidelberg},\ \bibinfo {year}
  {1985})\BibitemShut {NoStop}%
\bibitem [{\citenamefont {Le~Tacon}\ \emph {et~al.}(2011)\citenamefont
  {Le~Tacon}, \citenamefont {Ghiringhelli}, \citenamefont {Chaloupka},
  \citenamefont {Sala}, \citenamefont {Hinkov}, \citenamefont {Haverkort},
  \citenamefont {Minola}, \citenamefont {Bakr}, \citenamefont {Zhou},
  \citenamefont {Blanco-Canosa}, \citenamefont {Monney}, \citenamefont {Song},
  \citenamefont {Sun}, \citenamefont {Lin}, \citenamefont {De~Luca},
  \citenamefont {Salluzzo}, \citenamefont {Khaliullin}, \citenamefont
  {Schmitt}, \citenamefont {Braicovich},\ and\ \citenamefont
  {Keimer}}]{letacon2011}%
  \BibitemOpen
  \bibfield  {author} {\bibinfo {author} {\bibfnamefont {M.}~\bibnamefont
  {Le~Tacon}}, \bibinfo {author} {\bibfnamefont {G.}~\bibnamefont
  {Ghiringhelli}}, \bibinfo {author} {\bibfnamefont {J.}~\bibnamefont
  {Chaloupka}}, \bibinfo {author} {\bibfnamefont {M.~M.}\ \bibnamefont {Sala}},
  \bibinfo {author} {\bibfnamefont {V.}~\bibnamefont {Hinkov}}, \bibinfo
  {author} {\bibfnamefont {M.}~\bibnamefont {Haverkort}}, \bibinfo {author}
  {\bibfnamefont {M.}~\bibnamefont {Minola}}, \bibinfo {author} {\bibfnamefont
  {M.}~\bibnamefont {Bakr}}, \bibinfo {author} {\bibfnamefont {K.}~\bibnamefont
  {Zhou}}, \bibinfo {author} {\bibfnamefont {S.}~\bibnamefont {Blanco-Canosa}},
  \bibinfo {author} {\bibfnamefont {C.}~\bibnamefont {Monney}}, \bibinfo
  {author} {\bibfnamefont {Y.~T.}\ \bibnamefont {Song}}, \bibinfo {author}
  {\bibfnamefont {G.~L.}\ \bibnamefont {Sun}}, \bibinfo {author} {\bibfnamefont
  {C.~T.}\ \bibnamefont {Lin}}, \bibinfo {author} {\bibfnamefont {G.~M.}\
  \bibnamefont {De~Luca}}, \bibinfo {author} {\bibfnamefont {M.}~\bibnamefont
  {Salluzzo}}, \bibinfo {author} {\bibfnamefont {G.}~\bibnamefont
  {Khaliullin}}, \bibinfo {author} {\bibfnamefont {T.}~\bibnamefont {Schmitt}},
  \bibinfo {author} {\bibfnamefont {L.}~\bibnamefont {Braicovich}}, \ and\
  \bibinfo {author} {\bibfnamefont {B.}~\bibnamefont {Keimer}},\ }\href
  {http://www.nature.com/nphys/journal/v7/n9/abs/nphys2041.html} {\bibfield
  {journal} {\bibinfo  {journal} {Nature Physics}\ }\textbf {\bibinfo {volume}
  {7}},\ \bibinfo {pages} {725} (\bibinfo {year} {2011})}\BibitemShut {NoStop}%
\bibitem [{\citenamefont {Dean}\ \emph {et~al.}(2013)\citenamefont {Dean},
  \citenamefont {Dellea}, \citenamefont {Springell}, \citenamefont
  {Yakhou-Harris}, \citenamefont {Kummer}, \citenamefont {Brookes},
  \citenamefont {Liu}, \citenamefont {Sun}, \citenamefont {Strle},
  \citenamefont {Schmitt}, \citenamefont {Braicovich}, \citenamefont
  {Ghiringhelli}, \citenamefont {Božović},\ and\ \citenamefont
  {Hill}}]{dean2013}%
  \BibitemOpen
  \bibfield  {author} {\bibinfo {author} {\bibfnamefont {M.~P.~M.}\
  \bibnamefont {Dean}}, \bibinfo {author} {\bibfnamefont {G.}~\bibnamefont
  {Dellea}}, \bibinfo {author} {\bibfnamefont {R.~S.}\ \bibnamefont
  {Springell}}, \bibinfo {author} {\bibfnamefont {F.}~\bibnamefont
  {Yakhou-Harris}}, \bibinfo {author} {\bibfnamefont {K.}~\bibnamefont
  {Kummer}}, \bibinfo {author} {\bibfnamefont {N.~B.}\ \bibnamefont {Brookes}},
  \bibinfo {author} {\bibfnamefont {X.}~\bibnamefont {Liu}}, \bibinfo {author}
  {\bibfnamefont {Y.-J.}\ \bibnamefont {Sun}}, \bibinfo {author} {\bibfnamefont
  {J.}~\bibnamefont {Strle}}, \bibinfo {author} {\bibfnamefont
  {T.}~\bibnamefont {Schmitt}}, \bibinfo {author} {\bibfnamefont
  {L.}~\bibnamefont {Braicovich}}, \bibinfo {author} {\bibfnamefont
  {G.}~\bibnamefont {Ghiringhelli}}, \bibinfo {author} {\bibfnamefont
  {I.}~\bibnamefont {Božović}}, \ and\ \bibinfo {author} {\bibfnamefont
  {J.~P.}\ \bibnamefont {Hill}},\ }\href {\doibase 10.1038/nmat3723} {\bibfield
   {journal} {\bibinfo  {journal} {Nature Materials}\ }\textbf {\bibinfo
  {volume} {12}},\ \bibinfo {pages} {1019} (\bibinfo {year}
  {2013})}\BibitemShut {NoStop}%
\bibitem [{\citenamefont {Monney}\ \emph {et~al.}(2016)\citenamefont {Monney},
  \citenamefont {Schmitt}, \citenamefont {Matt}, \citenamefont {Mesot},
  \citenamefont {Strocov}, \citenamefont {Lipscombe}, \citenamefont {Hayden},\
  and\ \citenamefont {Chang}}]{monney2016}%
  \BibitemOpen
  \bibfield  {author} {\bibinfo {author} {\bibfnamefont {C.}~\bibnamefont
  {Monney}}, \bibinfo {author} {\bibfnamefont {T.}~\bibnamefont {Schmitt}},
  \bibinfo {author} {\bibfnamefont {C.~E.}\ \bibnamefont {Matt}}, \bibinfo
  {author} {\bibfnamefont {J.}~\bibnamefont {Mesot}}, \bibinfo {author}
  {\bibfnamefont {V.~N.}\ \bibnamefont {Strocov}}, \bibinfo {author}
  {\bibfnamefont {O.~J.}\ \bibnamefont {Lipscombe}}, \bibinfo {author}
  {\bibfnamefont {S.~M.}\ \bibnamefont {Hayden}}, \ and\ \bibinfo {author}
  {\bibfnamefont {J.}~\bibnamefont {Chang}},\ }\href {\doibase
  10.1103/PhysRevB.93.075103} {\bibfield  {journal} {\bibinfo  {journal} {Phys.
  Rev. B}\ }\textbf {\bibinfo {volume} {93}},\ \bibinfo {pages} {075103}
  (\bibinfo {year} {2016})}\BibitemShut {NoStop}%
\bibitem [{\citenamefont {Diallo}\ \emph {et~al.}(2009)\citenamefont {Diallo},
  \citenamefont {Antropov}, \citenamefont {Perring}, \citenamefont {Broholm},
  \citenamefont {Pulikkotil}, \citenamefont {Ni}, \citenamefont {Bud'ko},
  \citenamefont {Canfield}, \citenamefont {Kreyssig}, \citenamefont {Goldman},\
  and\ \citenamefont {McQueeney}}]{diallo2009}%
  \BibitemOpen
  \bibfield  {author} {\bibinfo {author} {\bibfnamefont {S.~O.}\ \bibnamefont
  {Diallo}}, \bibinfo {author} {\bibfnamefont {V.~P.}\ \bibnamefont
  {Antropov}}, \bibinfo {author} {\bibfnamefont {T.~G.}\ \bibnamefont
  {Perring}}, \bibinfo {author} {\bibfnamefont {C.}~\bibnamefont {Broholm}},
  \bibinfo {author} {\bibfnamefont {J.~J.}\ \bibnamefont {Pulikkotil}},
  \bibinfo {author} {\bibfnamefont {N.}~\bibnamefont {Ni}}, \bibinfo {author}
  {\bibfnamefont {S.~L.}\ \bibnamefont {Bud'ko}}, \bibinfo {author}
  {\bibfnamefont {P.~C.}\ \bibnamefont {Canfield}}, \bibinfo {author}
  {\bibfnamefont {A.}~\bibnamefont {Kreyssig}}, \bibinfo {author}
  {\bibfnamefont {A.~I.}\ \bibnamefont {Goldman}}, \ and\ \bibinfo {author}
  {\bibfnamefont {R.~J.}\ \bibnamefont {McQueeney}},\ }\href {\doibase
  10.1103/PhysRevLett.102.187206} {\bibfield  {journal} {\bibinfo  {journal}
  {Phys. Rev. Lett.}\ }\textbf {\bibinfo {volume} {102}},\ \bibinfo {pages}
  {187206} (\bibinfo {year} {2009})}\BibitemShut {NoStop}%
\bibitem [{\citenamefont {Zhao}\ \emph {et~al.}(2009)\citenamefont {Zhao},
  \citenamefont {Adroja}, \citenamefont {Yao}, \citenamefont {Bewley},
  \citenamefont {Li}, \citenamefont {Wang}, \citenamefont {Wu}, \citenamefont
  {Chen}, \citenamefont {Hu},\ and\ \citenamefont {Dai}}]{zhao2009}%
  \BibitemOpen
  \bibfield  {author} {\bibinfo {author} {\bibfnamefont {J.}~\bibnamefont
  {Zhao}}, \bibinfo {author} {\bibfnamefont {D.}~\bibnamefont {Adroja}},
  \bibinfo {author} {\bibfnamefont {D.-X.}\ \bibnamefont {Yao}}, \bibinfo
  {author} {\bibfnamefont {R.}~\bibnamefont {Bewley}}, \bibinfo {author}
  {\bibfnamefont {S.}~\bibnamefont {Li}}, \bibinfo {author} {\bibfnamefont
  {X.}~\bibnamefont {Wang}}, \bibinfo {author} {\bibfnamefont {G.}~\bibnamefont
  {Wu}}, \bibinfo {author} {\bibfnamefont {X.}~\bibnamefont {Chen}}, \bibinfo
  {author} {\bibfnamefont {J.}~\bibnamefont {Hu}}, \ and\ \bibinfo {author}
  {\bibfnamefont {P.}~\bibnamefont {Dai}},\ }\href
  {http://www.nature.com/nphys/journal/v5/n8/full/nphys1336.html} {\bibfield
  {journal} {\bibinfo  {journal} {Nature Physics}\ }\textbf {\bibinfo {volume}
  {5}},\ \bibinfo {pages} {555} (\bibinfo {year} {2009})}\BibitemShut {NoStop}%
\bibitem [{\citenamefont {Zhou}\ \emph {et~al.}(2013)\citenamefont {Zhou},
  \citenamefont {Huang}, \citenamefont {Monney}, \citenamefont {Dai},
  \citenamefont {Strocov}, \citenamefont {Wang}, \citenamefont {Chen},
  \citenamefont {Zhang}, \citenamefont {Dai}, \citenamefont {Patthey},
  \citenamefont {van~den Brink}, \citenamefont {Ding},\ and\ \citenamefont
  {Schmitt}}]{zhou2013}%
  \BibitemOpen
  \bibfield  {author} {\bibinfo {author} {\bibfnamefont {K.-J.}\ \bibnamefont
  {Zhou}}, \bibinfo {author} {\bibfnamefont {Y.-B.}\ \bibnamefont {Huang}},
  \bibinfo {author} {\bibfnamefont {C.}~\bibnamefont {Monney}}, \bibinfo
  {author} {\bibfnamefont {X.}~\bibnamefont {Dai}}, \bibinfo {author}
  {\bibfnamefont {V.~N.}\ \bibnamefont {Strocov}}, \bibinfo {author}
  {\bibfnamefont {N.-L.}\ \bibnamefont {Wang}}, \bibinfo {author}
  {\bibfnamefont {Z.-G.}\ \bibnamefont {Chen}}, \bibinfo {author}
  {\bibfnamefont {C.}~\bibnamefont {Zhang}}, \bibinfo {author} {\bibfnamefont
  {P.}~\bibnamefont {Dai}}, \bibinfo {author} {\bibfnamefont {L.}~\bibnamefont
  {Patthey}}, \bibinfo {author} {\bibfnamefont {J.}~\bibnamefont {van~den
  Brink}}, \bibinfo {author} {\bibfnamefont {H.}~\bibnamefont {Ding}}, \ and\
  \bibinfo {author} {\bibfnamefont {T.}~\bibnamefont {Schmitt}},\ }\href
  {http://www.nature.com/articles/ncomms2428} {\bibfield  {journal} {\bibinfo
  {journal} {Nature Communications}\ }\textbf {\bibinfo {volume} {4}},\
  \bibinfo {pages} {1470} (\bibinfo {year} {2013})}\BibitemShut {NoStop}%
\bibitem [{\citenamefont {Wang}\ \emph {et~al.}(2013)\citenamefont {Wang},
  \citenamefont {Zhang}, \citenamefont {Wang}, \citenamefont {Luo},
  \citenamefont {Regnault}, \citenamefont {Dai},\ and\ \citenamefont
  {Li}}]{wang2013}%
  \BibitemOpen
  \bibfield  {author} {\bibinfo {author} {\bibfnamefont {C.}~\bibnamefont
  {Wang}}, \bibinfo {author} {\bibfnamefont {R.}~\bibnamefont {Zhang}},
  \bibinfo {author} {\bibfnamefont {F.}~\bibnamefont {Wang}}, \bibinfo {author}
  {\bibfnamefont {H.}~\bibnamefont {Luo}}, \bibinfo {author} {\bibfnamefont
  {L.~P.}\ \bibnamefont {Regnault}}, \bibinfo {author} {\bibfnamefont
  {P.}~\bibnamefont {Dai}}, \ and\ \bibinfo {author} {\bibfnamefont
  {Y.}~\bibnamefont {Li}},\ }\href {\doibase 10.1103/PhysRevX.3.041036}
  {\bibfield  {journal} {\bibinfo  {journal} {Phys. Rev. X}\ }\textbf {\bibinfo
  {volume} {3}},\ \bibinfo {pages} {041036} (\bibinfo {year}
  {2013})}\BibitemShut {NoStop}%
\bibitem [{\citenamefont {Leong}\ \emph {et~al.}(2014)\citenamefont {Leong},
  \citenamefont {Lee}, \citenamefont {Lv},\ and\ \citenamefont
  {Phillips}}]{leong2014}%
  \BibitemOpen
  \bibfield  {author} {\bibinfo {author} {\bibfnamefont {Z.}~\bibnamefont
  {Leong}}, \bibinfo {author} {\bibfnamefont {W.-C.}\ \bibnamefont {Lee}},
  \bibinfo {author} {\bibfnamefont {W.}~\bibnamefont {Lv}}, \ and\ \bibinfo
  {author} {\bibfnamefont {P.}~\bibnamefont {Phillips}},\ }\href {\doibase
  10.1103/PhysRevB.90.125158} {\bibfield  {journal} {\bibinfo  {journal} {Phys.
  Rev. B}\ }\textbf {\bibinfo {volume} {90}},\ \bibinfo {pages} {125158}
  (\bibinfo {year} {2014})}\BibitemShut {NoStop}%
\bibitem [{\citenamefont {Dai}(2015)}]{dai2015}%
  \BibitemOpen
  \bibfield  {author} {\bibinfo {author} {\bibfnamefont {P.}~\bibnamefont
  {Dai}},\ }\href {\doibase 10.1103/RevModPhys.87.855} {\bibfield  {journal}
  {\bibinfo  {journal} {Rev. Mod. Phys.}\ }\textbf {\bibinfo {volume} {87}},\
  \bibinfo {pages} {855} (\bibinfo {year} {2015})}\BibitemShut {NoStop}%
\bibitem [{\citenamefont {Matsuhira}\ \emph {et~al.}(2011)\citenamefont
  {Matsuhira}, \citenamefont {Wakeshima}, \citenamefont {Hinatsu},\ and\
  \citenamefont {Takagi}}]{matsuhira2011}%
  \BibitemOpen
  \bibfield  {author} {\bibinfo {author} {\bibfnamefont {K.}~\bibnamefont
  {Matsuhira}}, \bibinfo {author} {\bibfnamefont {M.}~\bibnamefont
  {Wakeshima}}, \bibinfo {author} {\bibfnamefont {Y.}~\bibnamefont {Hinatsu}},
  \ and\ \bibinfo {author} {\bibfnamefont {S.}~\bibnamefont {Takagi}},\ }\href
  {http://dx.doi.org/10.1143/JPSJ.80.094701} {\bibfield  {journal} {\bibinfo
  {journal} {J. Phys. Soc. Jpn.}\ }\textbf {\bibinfo {volume} {80}},\ \bibinfo
  {pages} {094701} (\bibinfo {year} {2011})}\BibitemShut {NoStop}%
\bibitem [{\citenamefont {Nakayama}\ \emph {et~al.}(2016)\citenamefont
  {Nakayama}, \citenamefont {Kondo}, \citenamefont {Tian}, \citenamefont
  {Ishikawa}, \citenamefont {Halim}, \citenamefont {Bareille}, \citenamefont
  {Malaeb}, \citenamefont {Kuroda}, \citenamefont {Tomita}, \citenamefont
  {Ideta}, \citenamefont {Tanaka}, \citenamefont {Matsunami}, \citenamefont
  {Kimura}, \citenamefont {Inami}, \citenamefont {Ono}, \citenamefont
  {Kumigashira}, \citenamefont {Balents}, \citenamefont {Nakatsuji},\ and\
  \citenamefont {Shin}}]{nakayama2016}%
  \BibitemOpen
  \bibfield  {author} {\bibinfo {author} {\bibfnamefont {M.}~\bibnamefont
  {Nakayama}}, \bibinfo {author} {\bibfnamefont {T.}~\bibnamefont {Kondo}},
  \bibinfo {author} {\bibfnamefont {Z.}~\bibnamefont {Tian}}, \bibinfo {author}
  {\bibfnamefont {J.~J.}\ \bibnamefont {Ishikawa}}, \bibinfo {author}
  {\bibfnamefont {M.}~\bibnamefont {Halim}}, \bibinfo {author} {\bibfnamefont
  {C.}~\bibnamefont {Bareille}}, \bibinfo {author} {\bibfnamefont
  {W.}~\bibnamefont {Malaeb}}, \bibinfo {author} {\bibfnamefont
  {K.}~\bibnamefont {Kuroda}}, \bibinfo {author} {\bibfnamefont
  {T.}~\bibnamefont {Tomita}}, \bibinfo {author} {\bibfnamefont
  {S.}~\bibnamefont {Ideta}}, \bibinfo {author} {\bibfnamefont
  {K.}~\bibnamefont {Tanaka}}, \bibinfo {author} {\bibfnamefont
  {M.}~\bibnamefont {Matsunami}}, \bibinfo {author} {\bibfnamefont
  {S.}~\bibnamefont {Kimura}}, \bibinfo {author} {\bibfnamefont
  {N.}~\bibnamefont {Inami}}, \bibinfo {author} {\bibfnamefont
  {K.}~\bibnamefont {Ono}}, \bibinfo {author} {\bibfnamefont {H.}~\bibnamefont
  {Kumigashira}}, \bibinfo {author} {\bibfnamefont {L.}~\bibnamefont
  {Balents}}, \bibinfo {author} {\bibfnamefont {S.}~\bibnamefont {Nakatsuji}},
  \ and\ \bibinfo {author} {\bibfnamefont {S.}~\bibnamefont {Shin}},\ }\href
  {\doibase 10.1103/PhysRevLett.117.056403} {\bibfield  {journal} {\bibinfo
  {journal} {Phys. Rev. Lett.}\ }\textbf {\bibinfo {volume} {117}},\ \bibinfo
  {pages} {056403} (\bibinfo {year} {2016})}\BibitemShut {NoStop}%
\bibitem [{\citenamefont {Mandrus}\ \emph {et~al.}(2001)\citenamefont
  {Mandrus}, \citenamefont {Thompson}, \citenamefont {Gaal}, \citenamefont
  {Forro}, \citenamefont {Bryan}, \citenamefont {Chakoumakos}, \citenamefont
  {Woods}, \citenamefont {Sales}, \citenamefont {Fishman},\ and\ \citenamefont
  {Keppens}}]{mandrus2001}%
  \BibitemOpen
  \bibfield  {author} {\bibinfo {author} {\bibfnamefont {D.}~\bibnamefont
  {Mandrus}}, \bibinfo {author} {\bibfnamefont {J.~R.}\ \bibnamefont
  {Thompson}}, \bibinfo {author} {\bibfnamefont {R.}~\bibnamefont {Gaal}},
  \bibinfo {author} {\bibfnamefont {L.}~\bibnamefont {Forro}}, \bibinfo
  {author} {\bibfnamefont {J.~C.}\ \bibnamefont {Bryan}}, \bibinfo {author}
  {\bibfnamefont {B.~C.}\ \bibnamefont {Chakoumakos}}, \bibinfo {author}
  {\bibfnamefont {L.~M.}\ \bibnamefont {Woods}}, \bibinfo {author}
  {\bibfnamefont {B.~C.}\ \bibnamefont {Sales}}, \bibinfo {author}
  {\bibfnamefont {R.~S.}\ \bibnamefont {Fishman}}, \ and\ \bibinfo {author}
  {\bibfnamefont {V.}~\bibnamefont {Keppens}},\ }\href
  {http://journals.aps.org/prb/abstract/10.1103/PhysRevB.63.195104} {\bibfield
  {journal} {\bibinfo  {journal} {Phys. Rev. B}\ }\textbf {\bibinfo {volume}
  {63}},\ \bibinfo {pages} {195104} (\bibinfo {year} {2001})}\BibitemShut
  {NoStop}%
\bibitem [{\citenamefont {Padilla}\ \emph {et~al.}(2002)\citenamefont
  {Padilla}, \citenamefont {Mandrus},\ and\ \citenamefont
  {Basov}}]{padilla2002}%
  \BibitemOpen
  \bibfield  {author} {\bibinfo {author} {\bibfnamefont {W.~J.}\ \bibnamefont
  {Padilla}}, \bibinfo {author} {\bibfnamefont {D.}~\bibnamefont {Mandrus}}, \
  and\ \bibinfo {author} {\bibfnamefont {D.~N.}\ \bibnamefont {Basov}},\ }\href
  {http://journals.aps.org/prb/abstract/10.1103/PhysRevB.66.035120} {\bibfield
  {journal} {\bibinfo  {journal} {Phys. Rev. B}\ }\textbf {\bibinfo {volume}
  {66}},\ \bibinfo {pages} {035120} (\bibinfo {year} {2002})}\BibitemShut
  {NoStop}%
\bibitem [{\citenamefont {Yamaura}\ \emph {et~al.}(2012)\citenamefont
  {Yamaura}, \citenamefont {Ohgushi}, \citenamefont {Ohsumi}, \citenamefont
  {Hasegawa}, \citenamefont {Yamauchi}, \citenamefont {Sugimoto}, \citenamefont
  {Takeshita}, \citenamefont {Tokuda}, \citenamefont {Takata}, \citenamefont
  {Udagawa}, \citenamefont {Takigawa}, \citenamefont {Harima}, \citenamefont
  {Arima},\ and\ \citenamefont {Hiroi}}]{yamaura2012}%
  \BibitemOpen
  \bibfield  {author} {\bibinfo {author} {\bibfnamefont {J.}~\bibnamefont
  {Yamaura}}, \bibinfo {author} {\bibfnamefont {K.}~\bibnamefont {Ohgushi}},
  \bibinfo {author} {\bibfnamefont {H.}~\bibnamefont {Ohsumi}}, \bibinfo
  {author} {\bibfnamefont {T.}~\bibnamefont {Hasegawa}}, \bibinfo {author}
  {\bibfnamefont {I.}~\bibnamefont {Yamauchi}}, \bibinfo {author}
  {\bibfnamefont {K.}~\bibnamefont {Sugimoto}}, \bibinfo {author}
  {\bibfnamefont {S.}~\bibnamefont {Takeshita}}, \bibinfo {author}
  {\bibfnamefont {A.}~\bibnamefont {Tokuda}}, \bibinfo {author} {\bibfnamefont
  {M.}~\bibnamefont {Takata}}, \bibinfo {author} {\bibfnamefont
  {M.}~\bibnamefont {Udagawa}}, \bibinfo {author} {\bibfnamefont
  {M.}~\bibnamefont {Takigawa}}, \bibinfo {author} {\bibfnamefont
  {H.}~\bibnamefont {Harima}}, \bibinfo {author} {\bibfnamefont
  {T.}~\bibnamefont {Arima}}, \ and\ \bibinfo {author} {\bibfnamefont
  {Z.}~\bibnamefont {Hiroi}},\ }\href {\doibase 10.1103/PhysRevLett.108.247205}
  {\bibfield  {journal} {\bibinfo  {journal} {Phys. Rev. Lett.}\ }\textbf
  {\bibinfo {volume} {108}},\ \bibinfo {pages} {247205} (\bibinfo {year}
  {2012})}\BibitemShut {NoStop}%
\bibitem [{\citenamefont {Shi}\ \emph {et~al.}(2009)\citenamefont {Shi},
  \citenamefont {Guo}, \citenamefont {Yu}, \citenamefont {Arai}, \citenamefont
  {Belik}, \citenamefont {Sato}, \citenamefont {Yamaura}, \citenamefont
  {Takayama-Muromachi}, \citenamefont {Tian}, \citenamefont {Yang},
  \citenamefont {Li}, \citenamefont {Varga}, \citenamefont {Mitchell},\ and\
  \citenamefont {Okamoto}}]{shi2009}%
  \BibitemOpen
  \bibfield  {author} {\bibinfo {author} {\bibfnamefont {Y.~G.}\ \bibnamefont
  {Shi}}, \bibinfo {author} {\bibfnamefont {Y.~F.}\ \bibnamefont {Guo}},
  \bibinfo {author} {\bibfnamefont {S.}~\bibnamefont {Yu}}, \bibinfo {author}
  {\bibfnamefont {M.}~\bibnamefont {Arai}}, \bibinfo {author} {\bibfnamefont
  {A.~A.}\ \bibnamefont {Belik}}, \bibinfo {author} {\bibfnamefont
  {A.}~\bibnamefont {Sato}}, \bibinfo {author} {\bibfnamefont {K.}~\bibnamefont
  {Yamaura}}, \bibinfo {author} {\bibfnamefont {E.}~\bibnamefont
  {Takayama-Muromachi}}, \bibinfo {author} {\bibfnamefont {H.~F.}\ \bibnamefont
  {Tian}}, \bibinfo {author} {\bibfnamefont {H.~X.}\ \bibnamefont {Yang}},
  \bibinfo {author} {\bibfnamefont {J.~Q.}\ \bibnamefont {Li}}, \bibinfo
  {author} {\bibfnamefont {T.}~\bibnamefont {Varga}}, \bibinfo {author}
  {\bibfnamefont {J.~F.}\ \bibnamefont {Mitchell}}, \ and\ \bibinfo {author}
  {\bibfnamefont {S.}~\bibnamefont {Okamoto}},\ }\href {\doibase
  10.1103/PhysRevB.80.161104} {\bibfield  {journal} {\bibinfo  {journal} {Phys.
  Rev. B}\ }\textbf {\bibinfo {volume} {80}},\ \bibinfo {pages} {161104}
  (\bibinfo {year} {2009})}\BibitemShut {NoStop}%
\bibitem [{\citenamefont {Calder}\ \emph {et~al.}(2012)\citenamefont {Calder},
  \citenamefont {Garlea}, \citenamefont {McMorrow}, \citenamefont {Lumsden},
  \citenamefont {Stone}, \citenamefont {Lang}, \citenamefont {Kim},
  \citenamefont {Schlueter}, \citenamefont {Shi}, \citenamefont {Yamaura},
  \citenamefont {Sun}, \citenamefont {Tsujimoto},\ and\ \citenamefont
  {Christianson}}]{calder2012}%
  \BibitemOpen
  \bibfield  {author} {\bibinfo {author} {\bibfnamefont {S.}~\bibnamefont
  {Calder}}, \bibinfo {author} {\bibfnamefont {V.~O.}\ \bibnamefont {Garlea}},
  \bibinfo {author} {\bibfnamefont {D.~F.}\ \bibnamefont {McMorrow}}, \bibinfo
  {author} {\bibfnamefont {M.~D.}\ \bibnamefont {Lumsden}}, \bibinfo {author}
  {\bibfnamefont {M.~B.}\ \bibnamefont {Stone}}, \bibinfo {author}
  {\bibfnamefont {J.~C.}\ \bibnamefont {Lang}}, \bibinfo {author}
  {\bibfnamefont {J.-W.}\ \bibnamefont {Kim}}, \bibinfo {author} {\bibfnamefont
  {J.~A.}\ \bibnamefont {Schlueter}}, \bibinfo {author} {\bibfnamefont {Y.~G.}\
  \bibnamefont {Shi}}, \bibinfo {author} {\bibfnamefont {K.}~\bibnamefont
  {Yamaura}}, \bibinfo {author} {\bibfnamefont {Y.~S.}\ \bibnamefont {Sun}},
  \bibinfo {author} {\bibfnamefont {Y.}~\bibnamefont {Tsujimoto}}, \ and\
  \bibinfo {author} {\bibfnamefont {A.~D.}\ \bibnamefont {Christianson}},\
  }\href {http://journals.aps.org/prl/abstract/10.1103/PhysRevLett.108.257209}
  {\bibfield  {journal} {\bibinfo  {journal} {Phys. Rev. Lett}\ }\textbf
  {\bibinfo {volume} {108}},\ \bibinfo {pages} {257209} (\bibinfo {year}
  {2012})}\BibitemShut {NoStop}%
\bibitem [{\citenamefont {Du}\ \emph {et~al.}(2012)\citenamefont {Du},
  \citenamefont {Wan}, \citenamefont {Sheng}, \citenamefont {Dong},\ and\
  \citenamefont {Savrasov}}]{du2012}%
  \BibitemOpen
  \bibfield  {author} {\bibinfo {author} {\bibfnamefont {Y.}~\bibnamefont
  {Du}}, \bibinfo {author} {\bibfnamefont {X.}~\bibnamefont {Wan}}, \bibinfo
  {author} {\bibfnamefont {L.}~\bibnamefont {Sheng}}, \bibinfo {author}
  {\bibfnamefont {J.}~\bibnamefont {Dong}}, \ and\ \bibinfo {author}
  {\bibfnamefont {S.~Y.}\ \bibnamefont {Savrasov}},\ }\href {\doibase
  10.1103/PhysRevB.85.174424} {\bibfield  {journal} {\bibinfo  {journal} {Phys.
  Rev. B}\ }\textbf {\bibinfo {volume} {85}},\ \bibinfo {pages} {174424}
  (\bibinfo {year} {2012})}\BibitemShut {NoStop}%
\bibitem [{\citenamefont {Jung}\ \emph {et~al.}(2013)\citenamefont {Jung},
  \citenamefont {Song}, \citenamefont {Lee},\ and\ \citenamefont
  {Pickett}}]{jung2013}%
  \BibitemOpen
  \bibfield  {author} {\bibinfo {author} {\bibfnamefont {M.-C.}\ \bibnamefont
  {Jung}}, \bibinfo {author} {\bibfnamefont {Y.-J.}\ \bibnamefont {Song}},
  \bibinfo {author} {\bibfnamefont {K.-W.}\ \bibnamefont {Lee}}, \ and\
  \bibinfo {author} {\bibfnamefont {W.~E.}\ \bibnamefont {Pickett}},\ }\href
  {\doibase 10.1103/PhysRevB.87.115119} {\bibfield  {journal} {\bibinfo
  {journal} {Phys. Rev. B}\ }\textbf {\bibinfo {volume} {87}},\ \bibinfo
  {pages} {115119} (\bibinfo {year} {2013})}\BibitemShut {NoStop}%
\bibitem [{\citenamefont {Vecchio}\ \emph {et~al.}(2013)\citenamefont
  {Vecchio}, \citenamefont {Perucchi}, \citenamefont {Di~Pietro}, \citenamefont
  {Limaj}, \citenamefont {Schade}, \citenamefont {Sun}, \citenamefont {Arai},
  \citenamefont {Yamaura},\ and\ \citenamefont {Lupi}}]{lovecchio2013}%
  \BibitemOpen
  \bibfield  {author} {\bibinfo {author} {\bibfnamefont {I.~L.}\ \bibnamefont
  {Vecchio}}, \bibinfo {author} {\bibfnamefont {A.}~\bibnamefont {Perucchi}},
  \bibinfo {author} {\bibfnamefont {P.}~\bibnamefont {Di~Pietro}}, \bibinfo
  {author} {\bibfnamefont {O.}~\bibnamefont {Limaj}}, \bibinfo {author}
  {\bibfnamefont {U.}~\bibnamefont {Schade}}, \bibinfo {author} {\bibfnamefont
  {Y.}~\bibnamefont {Sun}}, \bibinfo {author} {\bibfnamefont {M.}~\bibnamefont
  {Arai}}, \bibinfo {author} {\bibfnamefont {K.}~\bibnamefont {Yamaura}}, \
  and\ \bibinfo {author} {\bibfnamefont {S.}~\bibnamefont {Lupi}},\ }\href
  {http://www.nature.com/articles/srep02990} {\bibfield  {journal} {\bibinfo
  {journal} {Scientific Reports}\ }\textbf {\bibinfo {volume} {3}},\ \bibinfo
  {pages} {2990} (\bibinfo {year} {2013})}\BibitemShut {NoStop}%
\bibitem [{\citenamefont {Kim}\ \emph {et~al.}(2016)\citenamefont {Kim},
  \citenamefont {Liu}, \citenamefont {Erg\"onenc}, \citenamefont {Toschi},
  \citenamefont {Khmelevskyi},\ and\ \citenamefont
  {Franchini}}]{bongjaekim2016}%
  \BibitemOpen
  \bibfield  {author} {\bibinfo {author} {\bibfnamefont {B.}~\bibnamefont
  {Kim}}, \bibinfo {author} {\bibfnamefont {P.}~\bibnamefont {Liu}}, \bibinfo
  {author} {\bibfnamefont {Z.}~\bibnamefont {Erg\"onenc}}, \bibinfo {author}
  {\bibfnamefont {A.}~\bibnamefont {Toschi}}, \bibinfo {author} {\bibfnamefont
  {S.}~\bibnamefont {Khmelevskyi}}, \ and\ \bibinfo {author} {\bibfnamefont
  {C.}~\bibnamefont {Franchini}},\ }\href {\doibase 10.1103/PhysRevB.94.241113}
  {\bibfield  {journal} {\bibinfo  {journal} {Phys. Rev. B}\ }\textbf {\bibinfo
  {volume} {94}},\ \bibinfo {pages} {241113} (\bibinfo {year}
  {2016})}\BibitemShut {NoStop}%
\bibitem [{\citenamefont {Calder}\ \emph {et~al.}(2015)\citenamefont {Calder},
  \citenamefont {Lee}, \citenamefont {Stone}, \citenamefont {Lumsden},
  \citenamefont {Lang}, \citenamefont {Feygenson}, \citenamefont {Zhao},
  \citenamefont {Yan}, \citenamefont {Shi}, \citenamefont {Sun}, \citenamefont
  {Tsujimoto}, \citenamefont {Yamaura},\ and\ \citenamefont
  {Christianson}}]{calder2015_naoso3}%
  \BibitemOpen
  \bibfield  {author} {\bibinfo {author} {\bibfnamefont {S.}~\bibnamefont
  {Calder}}, \bibinfo {author} {\bibfnamefont {J.~H.}\ \bibnamefont {Lee}},
  \bibinfo {author} {\bibfnamefont {M.~B.}\ \bibnamefont {Stone}}, \bibinfo
  {author} {\bibfnamefont {M.~D.}\ \bibnamefont {Lumsden}}, \bibinfo {author}
  {\bibfnamefont {J.~C.}\ \bibnamefont {Lang}}, \bibinfo {author}
  {\bibfnamefont {M.}~\bibnamefont {Feygenson}}, \bibinfo {author}
  {\bibfnamefont {Z.}~\bibnamefont {Zhao}}, \bibinfo {author} {\bibfnamefont
  {J.-Q.}\ \bibnamefont {Yan}}, \bibinfo {author} {\bibfnamefont {Y.~G.}\
  \bibnamefont {Shi}}, \bibinfo {author} {\bibfnamefont {Y.~S.}\ \bibnamefont
  {Sun}}, \bibinfo {author} {\bibfnamefont {Y.}~\bibnamefont {Tsujimoto}},
  \bibinfo {author} {\bibfnamefont {K.}~\bibnamefont {Yamaura}}, \ and\
  \bibinfo {author} {\bibfnamefont {A.~D.}\ \bibnamefont {Christianson}},\
  }\href {\doibase 10.1038/ncomms9916} {\bibfield  {journal} {\bibinfo
  {journal} {Nature Communications}\ }\textbf {\bibinfo {volume} {6}},\
  \bibinfo {pages} {9916} (\bibinfo {year} {2015})}\BibitemShut {NoStop}%
\bibitem [{\citenamefont {Calder}\ \emph {et~al.}(2017)\citenamefont {Calder},
  \citenamefont {Vale}, \citenamefont {Bogdanov}, \citenamefont {Donnerer},
  \citenamefont {Pincini}, \citenamefont {Moretti~Sala}, \citenamefont {Liu},
  \citenamefont {Upton}, \citenamefont {Casa}, \citenamefont {Shi},
  \citenamefont {Tsujimoto}, \citenamefont {Yamaura}, \citenamefont {Hill},
  \citenamefont {van~den Brink}, \citenamefont {McMorrow},\ and\ \citenamefont
  {Christianson}}]{calder2017_naoso3}%
  \BibitemOpen
  \bibfield  {author} {\bibinfo {author} {\bibfnamefont {S.}~\bibnamefont
  {Calder}}, \bibinfo {author} {\bibfnamefont {J.~G.}\ \bibnamefont {Vale}},
  \bibinfo {author} {\bibfnamefont {N.}~\bibnamefont {Bogdanov}}, \bibinfo
  {author} {\bibfnamefont {C.}~\bibnamefont {Donnerer}}, \bibinfo {author}
  {\bibfnamefont {D.}~\bibnamefont {Pincini}}, \bibinfo {author} {\bibfnamefont
  {M.}~\bibnamefont {Moretti~Sala}}, \bibinfo {author} {\bibfnamefont
  {X.}~\bibnamefont {Liu}}, \bibinfo {author} {\bibfnamefont {M.~H.}\
  \bibnamefont {Upton}}, \bibinfo {author} {\bibfnamefont {D.}~\bibnamefont
  {Casa}}, \bibinfo {author} {\bibfnamefont {Y.~G.}\ \bibnamefont {Shi}},
  \bibinfo {author} {\bibfnamefont {Y.}~\bibnamefont {Tsujimoto}}, \bibinfo
  {author} {\bibfnamefont {K.}~\bibnamefont {Yamaura}}, \bibinfo {author}
  {\bibfnamefont {J.~P.}\ \bibnamefont {Hill}}, \bibinfo {author}
  {\bibfnamefont {J.}~\bibnamefont {van~den Brink}}, \bibinfo {author}
  {\bibfnamefont {D.~F.}\ \bibnamefont {McMorrow}}, \ and\ \bibinfo {author}
  {\bibfnamefont {A.~D.}\ \bibnamefont {Christianson}},\ }\href {\doibase
  10.1103/PhysRevB.95.020413} {\bibfield  {journal} {\bibinfo  {journal} {Phys.
  Rev. B}\ }\textbf {\bibinfo {volume} {95}},\ \bibinfo {pages} {020413}
  (\bibinfo {year} {2017})}\BibitemShut {NoStop}%
\bibitem [{\citenamefont {Haverkort}(2010)}]{haverkort2010}%
  \BibitemOpen
  \bibfield  {author} {\bibinfo {author} {\bibfnamefont {M.~W.}\ \bibnamefont
  {Haverkort}},\ }\href {\doibase 10.1103/PhysRevLett.105.167404} {\bibfield
  {journal} {\bibinfo  {journal} {Phys. Rev. Lett.}\ }\textbf {\bibinfo
  {volume} {105}},\ \bibinfo {pages} {167404} (\bibinfo {year}
  {2010})}\BibitemShut {NoStop}%
\bibitem [{\citenamefont {Jia}\ \emph {et~al.}(2014)\citenamefont {Jia},
  \citenamefont {Nowadnick}, \citenamefont {Wohlfeld}, \citenamefont {Kung},
  \citenamefont {Chen}, \citenamefont {Johnston}, \citenamefont {Tohyama},
  \citenamefont {Moritz},\ and\ \citenamefont {Devereaux}}]{jia2014}%
  \BibitemOpen
  \bibfield  {author} {\bibinfo {author} {\bibfnamefont {C.}~\bibnamefont
  {Jia}}, \bibinfo {author} {\bibfnamefont {E.}~\bibnamefont {Nowadnick}},
  \bibinfo {author} {\bibfnamefont {K.}~\bibnamefont {Wohlfeld}}, \bibinfo
  {author} {\bibfnamefont {Y.}~\bibnamefont {Kung}}, \bibinfo {author}
  {\bibfnamefont {C.~C.}\ \bibnamefont {Chen}}, \bibinfo {author}
  {\bibfnamefont {S.}~\bibnamefont {Johnston}}, \bibinfo {author}
  {\bibfnamefont {T.}~\bibnamefont {Tohyama}}, \bibinfo {author} {\bibfnamefont
  {B.}~\bibnamefont {Moritz}}, \ and\ \bibinfo {author} {\bibfnamefont
  {T.}~\bibnamefont {Devereaux}},\ }\href
  {http://www.nature.com/articles/ncomms4314} {\bibfield  {journal} {\bibinfo
  {journal} {Nature Communications}\ }\textbf {\bibinfo {volume} {5}},\
  \bibinfo {pages} {3314} (\bibinfo {year} {2014})}\BibitemShut {NoStop}%
\bibitem [{\citenamefont {Kim}\ and\ \citenamefont
  {Khaliullin}(2017)}]{kim_khaliullin2017}%
  \BibitemOpen
  \bibfield  {author} {\bibinfo {author} {\bibfnamefont {B.~J.}\ \bibnamefont
  {Kim}}\ and\ \bibinfo {author} {\bibfnamefont {G.}~\bibnamefont
  {Khaliullin}},\ }\href {\doibase 10.1103/PhysRevB.96.085108} {\bibfield
  {journal} {\bibinfo  {journal} {Phys. Rev. B}\ }\textbf {\bibinfo {volume}
  {96}},\ \bibinfo {pages} {085108} (\bibinfo {year} {2017})}\BibitemShut
  {NoStop}%
\bibitem [{Note1()}]{Note1}%
  \BibitemOpen
  \bibinfo {note} {Data for other momentum transfers are presented in
  Ref.~\protect \rev@citealp {vale2018_prb}.}\BibitemShut {Stop}%
\bibitem [{\citenamefont {Fedders}\ and\ \citenamefont
  {Martin}(1966)}]{feddersmartin_1966}%
  \BibitemOpen
  \bibfield  {author} {\bibinfo {author} {\bibfnamefont {P.~A.}\ \bibnamefont
  {Fedders}}\ and\ \bibinfo {author} {\bibfnamefont {P.~C.}\ \bibnamefont
  {Martin}},\ }\href {\doibase 10.1103/PhysRev.143.245} {\bibfield  {journal}
  {\bibinfo  {journal} {Phys. Rev.}\ }\textbf {\bibinfo {volume} {143}},\
  \bibinfo {pages} {245} (\bibinfo {year} {1966})}\BibitemShut {NoStop}%
\bibitem [{\citenamefont {Inosov}\ \emph {et~al.}(2010)\citenamefont {Inosov},
  \citenamefont {Park}, \citenamefont {Bourges}, \citenamefont {Sun},
  \citenamefont {Sidis}, \citenamefont {Schneidewind}, \citenamefont {Hradil},
  \citenamefont {Haug}, \citenamefont {Lin}, \citenamefont {Keimer},\ and\
  \citenamefont {Hinkov}}]{inosov2010}%
  \BibitemOpen
  \bibfield  {author} {\bibinfo {author} {\bibfnamefont {D.}~\bibnamefont
  {Inosov}}, \bibinfo {author} {\bibfnamefont {J.}~\bibnamefont {Park}},
  \bibinfo {author} {\bibfnamefont {P.}~\bibnamefont {Bourges}}, \bibinfo
  {author} {\bibfnamefont {D.}~\bibnamefont {Sun}}, \bibinfo {author}
  {\bibfnamefont {Y.}~\bibnamefont {Sidis}}, \bibinfo {author} {\bibfnamefont
  {A.}~\bibnamefont {Schneidewind}}, \bibinfo {author} {\bibfnamefont
  {K.}~\bibnamefont {Hradil}}, \bibinfo {author} {\bibfnamefont
  {D.}~\bibnamefont {Haug}}, \bibinfo {author} {\bibfnamefont {C.}~\bibnamefont
  {Lin}}, \bibinfo {author} {\bibfnamefont {B.}~\bibnamefont {Keimer}}, \ and\
  \bibinfo {author} {\bibfnamefont {V.}~\bibnamefont {Hinkov}},\ }\href
  {http://www.nature.com/nphys/journal/v6/n3/abs/nphys1483.html} {\bibfield
  {journal} {\bibinfo  {journal} {Nature Physics}\ }\textbf {\bibinfo {volume}
  {6}},\ \bibinfo {pages} {178} (\bibinfo {year} {2010})}\BibitemShut {NoStop}%
\bibitem [{\citenamefont {Tucker}\ \emph {et~al.}(2014)\citenamefont {Tucker},
  \citenamefont {Fernandes}, \citenamefont {Pratt}, \citenamefont {Thaler},
  \citenamefont {Ni}, \citenamefont {Marty}, \citenamefont {Christianson},
  \citenamefont {Lumsden}, \citenamefont {Sales}, \citenamefont {Sefat},
  \citenamefont {Bud'ko}, \citenamefont {Canfield}, \citenamefont {Kreyssig},
  \citenamefont {Goldman},\ and\ \citenamefont {McQueeney}}]{tucker2014}%
  \BibitemOpen
  \bibfield  {author} {\bibinfo {author} {\bibfnamefont {G.~S.}\ \bibnamefont
  {Tucker}}, \bibinfo {author} {\bibfnamefont {R.~M.}\ \bibnamefont
  {Fernandes}}, \bibinfo {author} {\bibfnamefont {D.~K.}\ \bibnamefont
  {Pratt}}, \bibinfo {author} {\bibfnamefont {A.}~\bibnamefont {Thaler}},
  \bibinfo {author} {\bibfnamefont {N.}~\bibnamefont {Ni}}, \bibinfo {author}
  {\bibfnamefont {K.}~\bibnamefont {Marty}}, \bibinfo {author} {\bibfnamefont
  {A.~D.}\ \bibnamefont {Christianson}}, \bibinfo {author} {\bibfnamefont
  {M.~D.}\ \bibnamefont {Lumsden}}, \bibinfo {author} {\bibfnamefont {B.~C.}\
  \bibnamefont {Sales}}, \bibinfo {author} {\bibfnamefont {A.~S.}\ \bibnamefont
  {Sefat}}, \bibinfo {author} {\bibfnamefont {S.~L.}\ \bibnamefont {Bud'ko}},
  \bibinfo {author} {\bibfnamefont {P.~C.}\ \bibnamefont {Canfield}}, \bibinfo
  {author} {\bibfnamefont {A.}~\bibnamefont {Kreyssig}}, \bibinfo {author}
  {\bibfnamefont {A.~I.}\ \bibnamefont {Goldman}}, \ and\ \bibinfo {author}
  {\bibfnamefont {R.~J.}\ \bibnamefont {McQueeney}},\ }\href {\doibase
  10.1103/PhysRevB.89.180503} {\bibfield  {journal} {\bibinfo  {journal} {Phys.
  Rev. B}\ }\textbf {\bibinfo {volume} {89}},\ \bibinfo {pages} {180503}
  (\bibinfo {year} {2014})}\BibitemShut {NoStop}%
\bibitem [{\citenamefont {Diallo}\ \emph {et~al.}(2010)\citenamefont {Diallo},
  \citenamefont {Pratt}, \citenamefont {Fernandes}, \citenamefont {Tian},
  \citenamefont {Zarestky}, \citenamefont {Lumsden}, \citenamefont {Perring},
  \citenamefont {Broholm}, \citenamefont {Ni}, \citenamefont {Bud'ko},
  \citenamefont {Canfield}, \citenamefont {Li}, \citenamefont {Vaknin},
  \citenamefont {Kreyssig}, \citenamefont {Goldman},\ and\ \citenamefont
  {McQueeney}}]{diallo2010}%
  \BibitemOpen
  \bibfield  {author} {\bibinfo {author} {\bibfnamefont {S.~O.}\ \bibnamefont
  {Diallo}}, \bibinfo {author} {\bibfnamefont {D.~K.}\ \bibnamefont {Pratt}},
  \bibinfo {author} {\bibfnamefont {R.~M.}\ \bibnamefont {Fernandes}}, \bibinfo
  {author} {\bibfnamefont {W.}~\bibnamefont {Tian}}, \bibinfo {author}
  {\bibfnamefont {J.~L.}\ \bibnamefont {Zarestky}}, \bibinfo {author}
  {\bibfnamefont {M.}~\bibnamefont {Lumsden}}, \bibinfo {author} {\bibfnamefont
  {T.~G.}\ \bibnamefont {Perring}}, \bibinfo {author} {\bibfnamefont {C.~L.}\
  \bibnamefont {Broholm}}, \bibinfo {author} {\bibfnamefont {N.}~\bibnamefont
  {Ni}}, \bibinfo {author} {\bibfnamefont {S.~L.}\ \bibnamefont {Bud'ko}},
  \bibinfo {author} {\bibfnamefont {P.~C.}\ \bibnamefont {Canfield}}, \bibinfo
  {author} {\bibfnamefont {H.-F.}\ \bibnamefont {Li}}, \bibinfo {author}
  {\bibfnamefont {D.}~\bibnamefont {Vaknin}}, \bibinfo {author} {\bibfnamefont
  {A.}~\bibnamefont {Kreyssig}}, \bibinfo {author} {\bibfnamefont {A.~I.}\
  \bibnamefont {Goldman}}, \ and\ \bibinfo {author} {\bibfnamefont {R.~J.}\
  \bibnamefont {McQueeney}},\ }\href {\doibase 10.1103/PhysRevB.81.214407}
  {\bibfield  {journal} {\bibinfo  {journal} {Phys. Rev. B}\ }\textbf {\bibinfo
  {volume} {81}},\ \bibinfo {pages} {214407} (\bibinfo {year}
  {2010})}\BibitemShut {NoStop}%
\bibitem [{Note2()}]{Note2}%
  \BibitemOpen
  \bibinfo {note} {Further improvements in the quantitative agreement with the
  experimental data could likely be made by considering anisotropic spin
  fluctuations, or the effect of finite momentum resolution of the
  spectrometer. However, there are insufficient data to perform this
  robustly.}\BibitemShut {Stop}%
\end{thebibliography}
%

\end{document}